\documentclass{ws-ijmpd}
\usepackage[super,compress]{cite}
\begin{document}

\markboth{B. Elizaga Navascu\'es, M. Mart\'in-Benito, G.A. Mena Marug\'an}
{Hybrid Models in Loop Quantum Cosmology}

\catchline{}{}{}{}{}
%
%%%%%%%%%%%%%%%%%%%%%%%%%%%%%%%%%%%%%%%%%%%%%%%%%%%%%%%%%%%%%%%%%%%%

\title{Hybrid Models in Loop Quantum Cosmology}

\author{Beatriz Elizaga Navascu\'es}

\address{Instituto de Estructura de la Materia, IEM-CSIC, \\Serrano 121,
Madrid, 28006, Spain\\
beatriz.elizaga@iem.cfmac.csic.es}

\author{Mercedes Mart\'in-Benito}

\address{Radboud University Nijmegen, Institute for Mathematics, Astrophysics and Particle Physics,\\ Heyendaalseweg 135,
Nijmegen,  6525 AJ, The Netherlands\\
m.martin@hef.ru.nl}

\author{Guillermo A. Mena Marug\'an}

\address{Instituto de Estructura de la Materia, IEM-CSIC,\\ Serrano 121,
Madrid, 28006, Spain\\
mena@iem.cfmac.csic.es}

\maketitle

\begin{history}
\received{Day Month Year}
\revised{Day Month Year}
\end{history}

\begin{abstract}
In the framework of Loop Quantum Cosmology, inhomogeneous models are usually quantized by means of a hybrid approach that combines loop quantization techniques with standard quantum field theory methods. This approach is based on a splitting of the phase space in a homogeneous sector, formed by global, zero-modes, and an inhomogeneous sector, formed by the remaining, infinite number of modes, that describe the local degrees of freedom. Then, the hybrid quantization is attained by adopting a loop representation for the homogeneous gravitational sector, while a Fock representation is used for the inhomogeneities. The zero-mode of the Hamiltonian constraint operator couples the homogeneous and inhomogeneous sectors. The hybrid approach, therefore, is expected to provide a suitable quantum theory in regimes where the main quantum effects of the geometry are those affecting the zero-modes, while the inhomogeneities, still being quantum, can be treated in a more conventional way. This hybrid strategy was first proposed for the simplest cosmological midisuperspaces: the Gowdy models, and it has been later applied to the case of cosmological perturbations. This paper reviews the construction and main applications of hybrid Loop Quantum Cosmology. 
\end{abstract}


\ccode{PACS numbers: 04.60.Pp, 04.60.Kz, 98.80.Qc}

\section{Introduction}	

Since the pioneering works that founded Loop Quantum Cosmology (LQC) \cite{boj1a,boj1b,boj1c,boj1d,boj2,abl}, this field of research has undergone an impressive progress. Its application to the quantization of homogeneous cosmologies reveals that, while the behavior in semiclassical regions agrees with General Relativity, the loop quantization effects change drastically the dynamics around the Planck regime, making gravity repulsive. This behavior leads to the resolution of classical strong singularities, such as the Big Bang, since they get replaced by a quantum bounce \cite{aps} where energy densities reach a finite maximum critical value \cite{acs}. For details about the quantization of homogeneous models in LQC, we refer to the reviews \refcite{lqc1,lqc2,lqc3,lqc4} and references therein.

Despite the success in the quantization of homogeneous models, a realistic quantum cosmology applicable to our Universe asks for the introduction of inhomogeneities. 
%Indeed, it is an experimentally contrasted fact that cosmological perturbations played a prominent role in the Early Universe \cite{perturbations1,perturbations2}. They turned into classical density fluctuations that later, by gravitational instabilities, gave rise to the galaxies and other structures that we observe nowadays.
Inhomogeneous models are technically much more complicated than homogeneous ones because they possess an infinite number of degrees of freedom, or in other words, they are described by field theories. 
In order to face their quantization within the framework of LQC, a hybrid canonical approach has been proposed. It consists in quantizing the (global) zero-modes of the geometry following the LQC methods, while applying more conventional representation methods for the remaining degrees of freedom. This allows one to consistently deal with the field complexity of the system. More concretely, the zero-modes of the possibly existing matter fields are quantized \`a la Schr\"odinger, while the inhomogeneities of the geometry and of the matter fields (if present) are quantized \`a la Fock. 

The hybrid approach extends to inhomogeneous situations the procedure that was already adopted in homogeneous LQC with a (homogeneous as well) massless scalar field, for which only the geometry is quantized \`{a} la loop, while the scalar field is treated in a conventional Schr\"odinger representation \cite{aps}. In this homogeneous case, the discrete quantum nature that the geometry acquires owing to the loop quantization is enough to solve the singularity problem. The hybrid quantization of inhomogeneous models rests on a similar assumption, namely, that the zero-modes of the geometry already encode the main quantum geometry features, and then it is meaningful to apply a quantum-gravity inspired representation to them, while the rest of degrees of freedom can be treated in a conventional way. The hybrid approach ought to provide a suitable framework for a regime in between standard quantum field theory on curved spacetimes and the full loop quantum gravity (LQG) regime. The zero-mode of the Hamiltonian constraint couples the homogeneous and inhomogeneous sectors, and because of the loop quantization of the homogeneous gravitational sector, it retains the main quantum geometry effects. One then expects that physical states display singularity resolution, in the same way as it happens in the homogeneous case. 

The first model that was quantized following this hybrid LQC approach was the Gowdy model in vacuo, with three-torus ($T^{3}$) topology, and linear polarization \cite{hybrid-Gow1,hybrid-Gow2,hybrid-Gow3,hybrid-Gow4}. Gowdy models are midisuperspaces with compact spatial sections and two axial Killing fields \cite{gowdy1,gowdy2}. The model with the three-torus topology for the spatial sections and linearly polarized gravitational waves provides the simplest inhomogeneous cosmologies. After a partial gauge fixing, the reduced system can be regarded as gravitational waves propagating in one axial direction over a Bianchi I homogeneous background. Two global constraints remain in the model: a global momentum constraint on the inhomogeneities, and the zero-mode of the Hamiltonian constraint, so that the model is not completely deparameterized. This latter constraint is formed by the Hamiltonian constraint of the Bianchi I model, plus another term coupling the homogeneous and inhomogeneous sectors. Therefore, the hybrid quantization of this Gowdy model combines the quantization of the Bianchi I model in LQC \cite{awe} with a Fock quantization for the inhomogeneities of the gravitational field. As usual in field theories, there is an infinite ambiguity in the Fock representation chosen to quantize th
e field. Remarkably, previous studies on the Fock quantization of the completely deparameterized Gowdy model proved that two physically natural criteria select a unique (up to unitary equivalence) Fock representation for the gravitational waves \cite{men0,men1a,men1b,men2,men3}. In the completely deparameterized model there only remains a field propagating in a $2$-dimensional spacetime with circular spatial sections. The criteria needed to single out the representation are: a) invariance of the vacuum under the symmetries of the background where the field propagates, namely translations in the circle, and b) unitary implementability of the quantum dynamics. With this result at hand then, in the hybrid quantization, the same Fock representation as in the deparameterized model is chosen.  

The hybrid quantization of the $T^3$--Gowdy cosmology with linear polarization was later extended by including the presence of a massless scalar field with the same symmetries as the geometry \cite{gow-matter}. The main motivation to consider this model is that its homogeneous sector, the Bianchi I model with a homogeneous massless scalar, admits as isotropic solutions flat Friedmann-Robertson-Walker (FRW) cosmologies, that at certain scales give a good approximation to the dynamics of our Universe. In a convenient parameterization, the matter field contributes to the constraints in the very same way as the gravitational waves \cite{barbero}. As a consequence, the uniqueness results mentioned above for the Fock quantization \cite{men2,men3} apply to this nonvacuum case as well, and there is a preferred Fock description also for the inhomogeneities of the matter field. This model has served as a testbed to develop approximation methods to solve the resulting dynamics. These analyses show that the hybrid Gowdy model admits approximate solutions with a remarkable property: even if they are intrinsically inhomogeneous and anisotropic quantum states, they effectively satisfy the dynamics of a homogeneous and isotropic FRW model \cite{approx1}, possibly coupled to a perfect fluid \cite{approx2}, and with possible modifications to the geometry \cite{approx3}. The properties of the FRW geometry in LQC, specially the occurrence of the quantum bounce, are at the core of the approximations carried out when constructing such quantum states.

The hybrid approach is of course not limited to the Gowdy model, and has been applied as well to more realistic scenarios, such as inflationary FRW cosmologies with a scalar field and cosmological fluctuations. \cite{hyb-pert1,hyb-pert2,hyb-pert3,hyb-pert4,hyb-pert5}Unlike for the Gowdy model, where inhomogeneities are dealt with exactly, here the fluctuations are treated perturbatively and the action of the system is truncated at quadratic order in these perturbations. Again, the zero-mode of the Hamiltonian constraint couples the homogeneous sector with the inhomogeneities. Therefore, in this framework the fluctuations are a priori not treated as a test field propagating on a fixed background, but their back-reaction on the background is retained at the considered perturbative order in the action. The perturbed spacetime geometry is treated as a fully quantum entity. Remarkably, these analyses can be based on a covariant formulation of the system (at the perturbative order considered in the truncation), inasmuch as no gauge fixing needs to be adopted \cite{hyb-pert5}. Its hybrid quantization is obtained by employing an LQC representation for the degrees of freedom of the FRW geometry \cite{aps,mmo}, a Schr\"odinger representation for the zero-mode of the scalar field, and a Fock representation for the gauge-invariants that describe the inhomogeneous sector, that contains the Mukhanov-Sasaki (MS) field commonly used in standard cosmology.\cite{mukhanov}

As in the case of the Gowdy model, in principle there is an infinite ambiguity in the Fock representation chosen to quantize the gauge-invariant fluctuations. Luckily, for this case there exist as well uniqueness results fixing an equivalence class of Fock representations. These are again selected by the criteria of vacuum invariance under the spatial isometries of the background (that is considered to have compact spatial sections) and unitary implementability of the quantum dynamics\cite{uni0,uni1,uni2,uni3,uni4,uni7,uni8}. These results are obtained when the homogeneous sector is viewed as a fixed classical background, and the fluctuations as a test quantum field on it, namely, when back-reaction is neglected and only the fluctuations are quantized. This is the description employed in standard cosmology.\cite{mukhanov} Any element of the equivalence class of Fock representations selected in this way can be taken to provide the representation for the inhomogeneous sector in the hybrid quantization.

Within the hybrid scheme, one can analyze the influence of the quantization of the homogeneous sector on the dynamics of the perturbations at the level of truncation adopted in the action. This can be attained by applying a Born-Oppenheimer approximation, based on an ansatz for quantum states that separates the dependence on the homogeneous geometry from that on the inhomogeneous degrees of freedom, permitting the variation of both parts with respect to the zero-mode of the matter scalar field. \cite{hyb-pert4,hyb-pert5} In this way one can arrive at a scenario of quantum fields propagating on a quantum-corrected background, often called {\it dressed metric} approach that, under some additional assumptions, like e.g. the choice of a particular adiabatic vacuum,\footnote{The Fock representation associated with this vacuum is known to belong to the unique equivalence class selected by the criteria employed in the hybrid approach. \cite{uni8}} has allowed one to compute quantum gravity corrections to the primordial power spectra \cite{dres1,dres2,dres3,dres4,dres5}.

The structure of this work is the following. In Sec. \ref{gowdy} we present the hybrid quantization of the linearly polarized $T^3$--Gowdy model coupled to a massless scalar field. Sec. \ref{approx} is devoted to the construction of states that, in spite of being approximate solutions of the hybrid Gowdy model, they approximately satisfy as well an effective dynamics that corresponds to an FRW model. In Sec. \ref{perturbations} we discuss the application of the hybrid quantization to the case of models that are physically more relevant: perturbed FRW cosmologies coupled to a scalar field. Finally, in Sec. \ref{conclu} we summarize the main conclusions derived from hybrid LQC so far.

\section{Hybrid quantization of the linearly polarized $T^3$--Gowdy model}
\label{gowdy}

In order to present the main construction underlying the hybrid approach in LQC, we detail it for the simplest inhomogeneous cosmologies. These are Gowdy midisuperspaces, which are globally hyperbolic spacetimes, with compact spatial sections, and two axial Killing vectors \cite{gowdy1,gowdy2}. The case with three-torus topology for the spatial sections is the simplest one. We consider this model with linear polarization, namely the two Killing vectors are hypersurface orthogonal. In addition, we couple a massless scalar field with the same symmetries as the geometry. For further details one can consult Ref. \refcite{gow-matter}, that in turn builds on the construction for the vacuum model \cite{hybrid-Gow1,hybrid-Gow2,hybrid-Gow3,hybrid-Gow4}.

We choose coordinates $\{t,\theta,\sigma,\delta\}$ adapted to the symmetries, such that $\partial_\sigma$ and $\partial_\delta$ denote the Killing vectors. Then, the fields describing the model only depend on time and on the spatial coordinate $\theta \in S^1$. After performing a symmetry reduction and a partial gauge fixing \cite{hybrid-Gow2}, we obtain a reduced phase space formed by a pair of point-particle degrees of freedom, a gravitational field, and the massless scalar field. Using Fourier decomposition we split this phase space in two sectors. The mentioned point-particle degrees of freedom and the zero-modes of the remaining fields are global modes forming the \emph{homogeneous sector}.  It coincides with the phase space of the Bianchi I model coupled to a homogeneous massless scalar field $\phi$. The nonzero Fourier modes of both the gravitational and matter fields, $\xi$ and $\varphi$ respectively, form the \emph{inhomogeneous sector}. This reduced phase space is constrained by the zero-mode of the momentum constraint in the direction $\theta$, $\mathcal C_{\theta}$, that generates rigid rotations in the circle and only involves the inhomogeneous sector, and by the zero-mode of the Hamiltonian constraint, $\mathcal{C}_\text{G}=\mathcal{C}_{\text{hom}}+\mathcal{C}_{\text{inh}}$, formed by a homogeneous term $\mathcal{C}_{\text{hom}}$ that is the Hamiltonian constraint of the Bianchi I model, and by an additional term $\mathcal{C}_{\text{inh}}$ that couples the homogeneous and inhomogeneous sectors. Note that we do not deparameterize completely the system with the aim of imposing the remaining zero-mode of the Hamiltonian constraint at the quantum level. 

In order to quantize the homogeneous sector of the geometry by using LQC techniques, first we describe it in terms of the Ashtekar-Barbero variables of the Bianchi I model with three-torus topology. In an internal diagonal gauge these variables are given by the three components of the densitized triad $p_{j}$ and of the $su(2)$-connection $c_{j}$, with $j=\theta, \sigma,\delta$. They satisfy $\{c_{i}, p_{j}\}=8\pi \gamma G \delta_{ij}$, where $\gamma$ is the Immirzi parameter and $G$ is the Newton constant. Let us denote by $\mathcal{H}_{\text{kin}}^{\text{BI}}\otimes L^{2}(\mathbb{R},d\phi)$ the kinematical Hilbert space for the Bianchi I model in LQC \cite{awe}. Notice that for the zero-mode of the matter field, $\phi$, we choose a standard Schr\"odinger representation where the canonical conjugate momentum of $\phi$ acts as a derivative, $\hat{p}_\phi=-i\hbar\partial_\phi$, where $\hbar$ is the Planck constant. The construction of the geometry sector $\mathcal{H}_{\text{kin}}^{\text{BI}}$ mimics that of LQG in the sense that the connection is not defined in the quantum theory but only its holonomies. The inner product is discrete, so that the operators $\hat{p}_{j}$ have a point spectrum equal to the real line, and their mutual eigenstates $|p_\theta,p_\sigma,p_\delta\rangle$ form an orthonormal basis of $\mathcal{H}_{\text{kin}}^{\text{BI}}$. We denote the basic holonomy operators proposed in Ref. \refcite{awe} by $\hat{\mathcal{N}}_{\pm\bar{\mu}_{j}}$. Their action on the states $|p_\theta,p_\sigma,p_\delta\rangle$ is quite involved. It is convenient to relabel these states as $|\lambda_{\theta},\lambda_{\sigma},v\rangle$, where $\lambda_{j}\propto \text{sign}(p_{j})\sqrt{|p_{j}|}$ and $v=2\lambda_{\theta}\lambda_{\sigma}\lambda_{\delta}$ is proportional to the physical volume of the Bianchi I universe, or \emph{homogeneous} volume. The operators $\hat{\mathcal{N}}_{\pm\bar{\mu}_{j}}$ scale the label $\lambda_{j}$ in such a way that the label $v$ is simply shifted by one.\cite{awe}

In order to implement a Fock quantization of the inhomogeneous sector, we need to single out a preferred Fock representation. The completely deparameterized $T^3$--Gowdy model with linear polarization has been thoroughly studied and it has been proven that it admits a unique (up to unitary equivalence) Fock quantization \cite{men2,men3}. This result is quite relevant, inasmuch as it removes the freedom to choose among the infinite number of inequivalent Fock representations that may lead to different physics. Refs. \refcite{men2,men3} impose natural criteria to select a unique equivalence class of representations. These criteria consist in imposing vacuum invariance under translations in the circle, that is the symmetry generated by the only constraint present in the totally deparameterized model, $\mathcal C_{\theta}$, and in demanding that the dynamics can be implemented as a unitary operator in the quantum theory. Unitarity of the dynamics also imposes a concrete parameterization for the nonzero modes of both the gravitational field $\xi$ and the matter field $\varphi$. In this parameterization, both fields contribute to the system in the very same way. 

Then we represent the inhomogeneous sector of our hybrid model choosing the Fock space $\mathcal{F}^{\alpha}$ ($\alpha=\xi,\varphi$) of Refs.  \refcite{men2,men3}. An orthonormal basis is given by the \emph{n}-particle states $|\mathfrak{n^\alpha}\rangle=|\cdots, n_{-2}^{\alpha}, n_{-1}^{\alpha},n_1^{\alpha},n_2^{\alpha},\cdots\rangle $, where $n_m^{\alpha}$ denotes the occupation number of the field $\alpha$ in the mode $m\in\mathbb{Z}-\{0\}$. In addition, let $\hat{a}_{m}^{(\alpha)\dagger}$ and $\hat{a}_m^{(\alpha)}$ denote, respectively, the creation and annihilation operators. The total kinematical Hilbert space of the hybrid Gowdy model is thus $\mathcal{H}_{\text{kin}}=\mathcal{H}_{\text{kin}}^{\text{BI}}\otimes L^{2}(\mathbb{R},d\phi)\otimes \mathcal{F}^{\xi}\otimes \mathcal{F}^{\varphi}$, where $\mathcal{H}_{\text{kin}}^{\text{BI}}\otimes\mathcal{F}^{\xi}\otimes \mathcal{F}^{\varphi}$ is spanned by the basis states $|\lambda_{\theta},\lambda_{\sigma},v,\mathfrak{n}^\xi,\mathfrak{n}^\varphi\rangle$, and completed with
\begin{align}
\langle \lambda'_{\theta},\lambda'_{\sigma},v',\mathfrak{n'}^\xi,\mathfrak{n'}^\varphi|\lambda_{\theta},\lambda_{\sigma},v,\mathfrak{n}^\xi,\mathfrak{n}^\varphi\rangle=\delta_{\lambda'_{\theta},\lambda_{\theta}}\delta_{\lambda'_{\sigma},\lambda_{\sigma}}\delta_{v',v}\delta_{\mathfrak{n'}^\xi,\mathfrak{n}^\xi}\delta_{\mathfrak{n'}^\varphi,\mathfrak{n}^\varphi}
\end{align} 
as the inner product, where $\delta_{x',x}$ denotes the Kronecker delta. We can now represent the constraints as operators, densely defined on this Hilbert space. Choosing normal ordering, the generator of the translations in the circle reads \cite{gow-matter}
\begin{align}
\widehat{\mathcal{C}}_{\theta}=\sum_{m=1}^{\infty}m\left(\widehat{X}^{\xi}_{m}+\widehat{X}^{\varphi}_{m}\right)\; , \qquad\widehat{X}_{m}^{\alpha}=\hat{a}^{(\alpha)\dagger}_{m}\hat{a}^{(\alpha)}_{m}-\hat{a}^{(\alpha)\dagger}_{-m}\hat{a}^{(\alpha)}_{-m}\,.
\end{align}
The imposition of this constraint on $\mathcal{F}^{\xi}\otimes\mathcal{F}^{\varphi}$ leads to the condition,
\begin{align}\label{mom}
\sum_{m=1}^{\infty}m(X^{\xi}_{m}+ X^{\varphi}_{m})=0\;, \qquad X^{\alpha}_{m}=n^{\alpha}_{m}-n^{\alpha}_{-m}\,.
\end{align}
The $n$--particle states that satisfy this condition provide a proper Fock subspace $\mathcal{F_{\text{p}}}\subset \mathcal{F}^{\xi}\otimes\mathcal{F}^{\varphi}$ that is unitarily equivalent to the physical Fock space of the inhomogeneities in the deparameterized model, obtained as in the vacuum case \cite{men2,men3}.

In constructing the quantum Hamiltonian constraint, for the operators acting nontrivially on the inhomogeneous sector, we choose, e.g., normal ordering again.\cite{approx1} For the operators acting on the homogeneous sector we choose a convenient symmetrization \cite{hybrid-Gow3}. As a result, the Hamiltonian constraint operator decouples the states of zero homogeneous volume $v$. Consequently, we can remove from our kinematical theory the states which are the analog of the classical singularity (those with vanishing $v$). Besides, the action of this constraint turns out not to relate states with different signs of the variables $v$, $\lambda_{\theta}$, and $\lambda_{\sigma}$. Hence we can restrict the study to the sector with, e.g., strictly positive labels for these homogeneous geometry variables. For later convenience we define $\Lambda_j=\ln(\lambda_j)$.

The resulting Hamiltonian constraint reads $\widehat{\mathcal{C}}_\text{G}=\widehat{\mathcal{C}}_\text{hom}+\widehat{\mathcal{C}}_\text{inh}$, where \cite{hybrid-Gow3,hybrid-Gow4}
\begin{align}\label{CBI}
\widehat{\mathcal{C}}_\text{hom}&=-\dfrac{\pi G \hbar^2} {16}\!\!\sum_{i\neq j}\!\sum_{j}\!\widehat{\Theta}_{i}\widehat{\Theta}_{j}+\dfrac{\hat{p}_\phi^{2}}{2}\;,\\ \label{inh} \widehat{\mathcal{C}}_\text{inh}&=\dfrac{2\pi G\hbar^2}{\beta}\widehat{e^{2\Lambda_\theta}}\widehat{H}_{0}+\dfrac{\pi G \hbar^2\beta}{16}\widehat{e^{-2\Lambda_\theta}}\widehat{D}(\widehat{\Theta}_{\delta}\!+\!\widehat{\Theta}_{\sigma})^{2}\widehat{D}\widehat{H}_{I}\;.
\end{align}
Here $i,j \in \{\theta,\sigma,\delta\}$. As mentioned before, the first term is the Hamiltonian constraint operator in LQC of the Bianchi I model with a homogeneous massless scalar field. We have introduced the quantum version of ${c_{j}p_{j}}$, given by $\pi G \hbar \gamma \widehat{\Theta}_{j}$ with
\begin{align}
\widehat{\Theta}_{j}=\frac{1}{2i}\widehat{\sqrt{|v|}}\left[\left(\hat{\mathcal{N}}_{2\bar{\mu}_{j}}-\hat{\mathcal{N}}_{-2\bar{\mu}_{j}}\right)\widehat{\text{sign}(p_{j})}+\widehat{{\text{sign}}(p_{j})}\left(\hat{\mathcal{N}}_{2\bar{\mu}_{j}}-\hat{\mathcal{N}}_{-2\bar{\mu}_{j}}\right)\right]\widehat{\sqrt{|v|}}\;.
\end{align}
In Eq. \eqref{inh}, $\beta$ is a constant related with parameters of the loop quantization.\cite{gow-matter} The operator $\widehat{D}$ represents the product of the volume by its inverse (which is regularized in a standard way in LQC \cite{abl}), and its action is $\widehat{D}|v\rangle=v\left(\sqrt{|v+1|}-\sqrt{|v-1|}\right)^2|v\rangle$. The operators $\widehat{H}_{\text{0}}$ and $\widehat{H}_{I}$ in the inhomogeneous term are given by
\begin{align}
\widehat{H}_{0}=\sum_{\alpha}\sum_{m=1}^{\infty}m \widehat{N}^{\alpha}_{m}\;, \quad \widehat{H}_{I}=\sum_{\alpha} \sum_{m=1}^{\infty}\dfrac{1}{m}\left(\widehat{N}^{\alpha}_{m}+ \hat{a}^{(\alpha)\dagger}_{m} \hat{a}^{(\alpha)\dagger}_{-m} + \hat{a}^{(\alpha)}_{m} \hat{a}^{(\alpha)}_{-m}\right)\;,
\end{align}
where $\alpha \in \{\xi,\varphi\}$ again, and $\widehat{N}^{\alpha}_{m}=\hat{a}^{(\alpha)\dagger}_{m} \hat{a}^{(\alpha)}_{m} + \hat{a}^{(\alpha)\dagger}_{-m} \hat{a}^{(\alpha)}_{-m}$. Note that the inhomogeneities of both fields contribute to the constraints in exactly the same way. 

The Hamiltonian constraint operator $\widehat{\mathcal{C}}_\text{G}$ does not relate all states with different values of $v$ and $\Lambda_j$ $(j= \theta,\sigma)$, but \emph{superselects} different sectors. The superselection sectors in $v$ are semilattices of step four $\mathcal{L}_{\varepsilon}^+=\{\varepsilon+4k, k\in\mathbb{N}\}$ determined by the initial point $\varepsilon\in(0,4]$. The superselection sectors in $\Lambda_j$ are more involved. Given initial data $\Lambda_{j}^{\ast}$ and $\varepsilon$, the corresponding values of $\Lambda_j$ are of the form $\Lambda_j=\Lambda_{j}^{\ast}+\Lambda_{\varepsilon}$, where $\Lambda_\varepsilon$ belongs to a certain set $\mathcal{W}_\varepsilon$ that is countable and dense in $\mathbb{R}$ \cite{hybrid-Gow3}. 

The involved form of $\widehat{\Theta}_{j}$ complicates the proof of its self-adjointness. It is then common to assume that $\widehat{\mathcal{C}}_\text{hom}$ is self-adjoint \cite{awe}, as well as $\widehat{\mathcal{C}}_\text{G}$ \cite{hybrid-Gow3,hybrid-Gow4}. Regardless of this assumption, one can formally analyze the solutions of the Gowdy model, which
% The quantum Hamiltonian constraint leads to a difference equation in the variable $v$, and then we can regard it as an evolution equation in $v$.  The solutions 
are completely determined by the data on the initial $v$-section, $v=\varepsilon$. This property allows one to characterize the physical Hilbert space as  the Hilbert space of these initial data, whose inner product is determined by imposing reality conditions in a complete set of observables. The result is
$\mathcal{H}_{\text{phys}}=\mathcal{H}_{\text{phys}}^{\text{BI}}\otimes L^{2}(\mathbb{R},d\phi) \otimes \mathcal{F}_{\text{p}}$, where $\mathcal{H}_{\text{phys}}^{\text{BI}}$ is the physical Hilbert space for Bianchi I cosmologies given in Ref. \refcite{hybrid-Gow4}.

\section{Approximation methods: modelling effective FRW cosmologies from states of the Gowdy model}
\label{approx}

Even if the hybrid quantization provides a well-defined operator for the Hamiltonian constraint (as we have just seen in the example of the Gowdy model), solving exactly this constraint is a highly difficult task, if not impossible. In order to get insight into the properties of physical states, approximation methods are needed. The $T^3$--Gowdy model with linear polarization serves again as a suitable testbed to develop such approximations. In order to simplify this discussion, we consider a simpler subsystem of the model built in Sec. \ref{gowdy}. We impose that the homogeneous sector displays local rotational symmetry (LRS), so that $c_\sigma=c_\delta$, $p_\sigma=p_\delta$. Note that this symmetry is viable because the Gowdy model is symmetric under the interchange $\sigma \leftrightarrow \delta$.  At the quantum level, we implement LRS via the map \cite{gow-matter}
\begin{align}\label{lrs}
|\Psi(\Lambda_{\theta},\Lambda_{\sigma}, v)\rangle \qquad \longrightarrow \qquad \sum_{\Lambda_{\sigma}}|\Psi(\Lambda_{\theta},\Lambda_{\sigma}, v)\rangle \equiv|\psi(\Lambda_{\theta},v)\rangle\;,
\end{align}
from Gowdy states (spanned over the homogeneous geometry by the basis of Bianchi I states $|\Lambda_{\theta},\Lambda_{\sigma}, v\rangle$) to LRS-Gowdy states (spanned by the basis of states $|\Lambda_{\theta},v\rangle$). Here the sum in $\Lambda_{\sigma}$ is carried out over the corresponding superselection sector.

After this LRS reduction, we get $\widehat{\Theta}_\sigma=\widehat{\Theta}_\delta=:\widehat{\Omega}$, which is essentially self-adjoint. \cite{mmo} Introducing the operator $\widehat{\Theta}= \widehat{\Theta}_\theta-\widehat{\Omega}$ we can write the Bianchi I term \eqref{CBI} as the Hamiltonian constraint in LQC of the flat FRW model coupled to a massless scalar, $\widehat{\mathcal{C}}_{\text{FRW}}$, plus a contribution that accounts for the anisotropies:
\begin{align}\label{cfrw}
\widehat{\mathcal{C}}^\text{LRS}_\text{hom}&= \widehat{\mathcal{C}}_{\text{FRW}}-\frac{\pi G\hbar^2}{8}(\widehat{\Omega}\widehat{\Theta}+\widehat{\Theta}\widehat{\Omega})\;,\qquad  \widehat{\mathcal{C}}_{\text{FRW}}=-\frac{3\pi G\hbar^2}{8}\widehat{\Omega}^2+\frac{\hat{p}_\phi^2}{2}\;.
\end{align}
On the other hand, the inhomogeneous term \eqref{inh} reduces to
\begin{align}
\widehat{\mathcal{C}}^\text{LRS}_\text{inh}&=\dfrac{2\pi G\hbar^2}{\beta}\widehat{e^{2\Lambda_\theta}}\widehat{H}_{0}+\dfrac{\pi G \hbar^2\beta}{4}\widehat{e^{-2\Lambda_\theta}}\widehat{D}
\widehat{\Omega}^{2}\widehat{D}\widehat{H}_{I}\;.\end{align}

Ref. \refcite{approx3} provides approximate solutions to the Gowdy Hamiltonian constraint $\widehat{\mathcal C}^\text{LRS}_\text{G}=\widehat{\mathcal{C}}^\text{LRS}_\text{hom}+\widehat{\mathcal{C}}^\text{LRS}_\text{inh}$ such that they are as well approximate solutions to the simpler constraint 
\begin{align}\label{const-app-prime}
\widehat{\mathcal C}_{\text{app}}=\widehat{\mathcal C}_{\text{FRW}}+\frac{2\pi G \hbar^{2}}{\beta}e^{2\bar\Lambda(\hat{\omega})}\widehat{H}_0=-\frac{3\pi G\hbar^2}{8}\widehat{\Omega}^2+\frac{\hat{p}_\phi^2}{2}+\frac{2\pi G\hbar^{2}}{\beta} e^{2\bar\Lambda(\hat{\omega})}\widehat{H}_0\;.
\end{align}
Here $\hat{\omega}$ is any self-adjoint operator defined on the homogeneous and isotropic geometry part of the kinematical Hilbert space (that is, the space spanned by the states $| v\rangle$). For instance, $\hat{\omega}$ may be just a constant, case studied in Ref. \refcite{approx1}, or equal to $\hat{v}$, as studied in Ref. \refcite{approx2}. On the other hand, the function $\bar\Lambda(\omega)$ is arbitrary except for two conditions. It has to be smooth and much larger than the unit for all values of $\omega$, in the spectrum of $\hat{\omega}$, that contribute significantly to the support of the solutions. 

To illustrate how those solutions look like, let us focus on the case $\hat{\omega}=\hat{v}$. Then, those states are characterized by profiles of the form \cite{approx2}
\begin{align}\label{profiles}
\Psi(\Lambda_\theta,v,\phi,\mathfrak{n}^{\xi},\mathfrak{n}^{\varphi})&= \int_{-\infty}^{\infty} dp_{\phi}\, e_{p_\phi}(\phi) N(v,p_\phi,\mathfrak{n}^{\xi},\mathfrak{n}^{\varphi}) f(\Lambda_\theta,v)\; ,
\end{align}
where $e_{p_\phi}(\phi)=e^{\frac{i}{\hbar}p_\phi\phi}/\sqrt{2\pi\hbar}$ are the plane-waves that diagonalize $\hat{p}_\phi^2=-\hbar^2\partial^2_\phi$. The rest of objects in Eq. \eqref{profiles} verify the following properties:
\begin{itemlist}
\item
The dependence of the function $N(v,p_\phi,\mathfrak{n}^{\xi},\mathfrak{n}^{\varphi})$ on the occupancy numbers of the $n$--particle states must be chosen in such a way that the momentum constraint \eqref{mom} is satisfied, and it is assumed that the content of inhomogeneities is not large.\cite{approx2}
\item 
The function $N(v,p_\phi,\mathfrak{n}^{\xi},\mathfrak{n}^{\varphi})$ is supported only on a region of sufficiently large values of $p_\phi$.\cite{approx2}
\item $N(v,p_\phi,\mathfrak{n}^{\xi},\mathfrak{n}^{\varphi})$ has to be highly suppressed for $v\lesssim v_m$, with $v_m\in \mathcal{L}^+_\varepsilon$ a certain value of the volume such that $v_m\gg 10$.
\item The function $f(\Lambda_\theta,v)$ has the form 
%$f(\Lambda_\theta,v)= \exp({-\frac{\sigma^{2}_{s}}{2q_{\epsilon}^{2}}[\Lambda_\theta-\bar{\Lambda}(v)]^{2}})$.
\begin{align}\label{ani-omega}
f(\Lambda_\theta,v)= e^{-\frac{\sigma^{2}_{s}}{2q_{\epsilon}^{2}}[\Lambda_\theta-\bar{\Lambda}(v)]^{2}}.
\end{align}
Here, $q_\varepsilon=\ln(1+2/v_m)$ and $q_\varepsilon \ll \sigma_s \ll 1$. Furthermore, the peak of this Gaussian must verify that $\bar\Lambda(v\pm4)\simeq\bar\Lambda(v)\gg 1$ for all $v$ in the support of $N(v,p_\phi,\mathfrak{n}^{\xi},\mathfrak{n}^{\varphi})$. 
\end{itemlist}
Therefore, these states are characterized by a Gaussian profile in the anisotropy variable $\Lambda_\theta$, peaked on a large value of it. Their properties make it possible to disregard the action of the anisotropy operator $(\widehat{\Omega}\widehat{\Theta}+\widehat{\Theta}\widehat{\Omega})$ present in $\widehat{\mathcal{C}}^\text{LRS}_\text{hom}$, and of the second term of $\widehat{\mathcal{C}}^\text{LRS}_\text{inh}$ when acting on these states. It is also possible to approximate the action of $\widehat{e^{2\Lambda_\theta}}$, present in the first term of $\widehat{\mathcal{C}}^\text{LRS}_\text{inh}$, by the action of the operator ${e^{2\bar\Lambda(\hat{v})}}$. Ref. \refcite{approx2} contains the explicit construction of states with the mentioned properties, and shows that if 
\begin{align} 
\bar\Lambda(v) =
\begin{cases}
\ln \left[v_0^{(1-w)/2}\right], & \text{if }v\leq v_0 \\ \ln \left[v^{(1-w)/2}\right], & \text{if }v>v_0
\end{cases}
\end{align}
with $v_0\gg \exp {\{2}/{[(1-w)]}\}$ and $w<1$ two constants, then the resulting states mimic the behavior of a perfect fluid with constant pressure-to-density ratio equal to $w$. This statement has to be understood in the sense that the states are approximate solutions of a flat FRW model coupled to such a perfect fluid, with Hamiltonian
\begin{align}\label{const-appf}
\widehat{\mathcal C}_{\text{FRW+PF}}=-\frac{3\pi G\hbar^2}{8}\widehat{\Omega}^2+\frac{\hat{p}_\phi^2}{2}+\alpha(1-w) \hat{v}^{1-w}\;.
\end{align}
Here $\alpha$ is a constant related to $\beta$ and to $H_{0}(\mathfrak{n}^{\xi},\mathfrak{n}^{\varphi})=\sum_{m\in\mathbb{Z}-\{0\}}|m|(n^\xi_m+n^\varphi_m)$.

Let us notice that such an effective description with a coupling to one of those perfect fluids with $w<1$ begins only when the evolution reaches the volume $v_{0}$, while we find an equation of state of a homogeneous massless scalar field for smaller values of $v$. The phase with $w<1$ holds then indefinitely for $v>v_{0}$ by the very construction of the states. The case $w=1$, realized when $\bar\Lambda$ is constant, corresponds to a massless scalar field, and is discussed in Ref. \refcite{approx1}. Moreover, the generalization done in Ref. \refcite{approx3}, with more generic $\bar\Lambda(\hat{\omega})$, accounts for states that mimic the behavior of an FRW model coupled to several perfect fluids, and with modifications to the FRW geometry similar to those considered in modified theories of gravity.

\section{Hybrid quantization of gauge-invariant cosmological fluctuations}
\label{perturbations}

Nowadays, precision cosmology is undergoing an outstanding progress \cite{precision1,precision2}. The latest cosmological observations provide highly accurate data, that may open a window to measure quantum geometry effects of the Early Universe. In this context, a hope of the quantum gravity community is to develop a quantum cosmology formalism capable of leading to testable predictions. 

The best established framework that conciliates the theoretical models of the Early Universe with observations is the theory of cosmological perturbations starting in an inflationary FRW scenario \cite{mukhanov}. The standard analyses study the perturbations during inflation within the scheme of quantum field theory in a classical and fixed curved spacetime. Despite the success of this treatment, the challenge for quantum cosmology is to build a formalism which includes simultaneously both the quantum geometry and the perturbations, with interplay between them. The aim is to elucidate whether the relics of the quantum fluctuations of the Early Universe may encode information about the quantum character of the spacetime geometry itself. 

Hybrid LQC provides a suitable framework to address this question. Indeed, it has been already applied to the quantization of cosmological perturbations around FRW spacetimes minimally coupled to a scalar field \cite{hyb-pert1,hyb-pert2,hyb-pert3,hyb-pert4,hyb-pert5}. These studies truncate the action at quadratic order in perturbations, and employ a decomposition of the fields in modes constructed out of the eigenfunctions of the Laplacian, defined on the spatial sections of the homogeneous and isotropic unperturbed model  \cite{HH}. The perturbations are then the nonzero modes in that decomposition, and form the inhomogeneous sector, while the zero-modes provide the homogeneous sector. The truncated system is subject to the zero-mode of the Hamiltonian constraint, and also to inhomogeneous constraints that are linear in perturbations. They arise from the perturbation of the Hamiltonian and spatial diffeomorphisms constraints around the FRW geometry in General Relativity. 

In the following we summarize these studies, in particular that of Ref. \refcite{hyb-pert5}, which, as main novel result, introduces a formulation of the classical system specially designed to preserve covariance (within the considered truncation),  inasmuch as no gauge fixing is adopted. The analysis is particularized to flat compact spatial sections and to scalar perturbations, although the formalism can be straightforwardly extended to other compact topologies and to tensor perturbations. In this covariant formulation the perturbations are described in terms of MS gauge-invariants \cite{mukhanov} and linear perturbative constraints, together with the variables canonically conjugate to them. This set is completed into a canonical one for the entire system, including the homogeneous degrees of freedom, so that homogeneous and inhomogenous sectors form together a symplectic manifold. The zero-mode of the Hamiltonian constraint of this truncated system is formed by the contribution of the homogeneous sector (that of the unperturbed flat FRW model with a homogeneous scalar field) plus a term quadratic in the perturbations. It retains the back-reaction on the homogeneous background up to the considered truncation order in perturbations. In this covariant formulation, the algebra of constraints under Poisson brackets is abelianized (at the perturbative order considered).

Let us introduce some notation for concreteness. In the flat unperturbed model with compact spatial sections, real Fourier modes are eigenmodes of the Laplacian and provide a basis to expand the considered scalar perturbations. These modes are labelled by any tuple of integers $\vec n=(n_1,n_2,n_3)\in\mathbb Z^3$ such that its first nonvanishing component is strictly positive, and by an extra index $\epsilon=\{+,-\}$ that distinguishes between sine and cosine functions. We denote by $-\omega_n^2 \propto -\vec n\cdot\vec n$ the corresponding Laplacian eigenvalues.\cite{hyb-pert5} In the expansion of the inhomogeneities, the vanishing tuple $\vec n$ is not included, as this mode is accounted for in the homogeneous sector. Let $\{V^{\vec n,\epsilon}_{p_l}\}\equiv\{\pi_{v_{\vec n,\epsilon}}, \mathcal C^{\vec n,\epsilon}_{|1}, \mathcal C^{\vec n,\epsilon}_{\_1}\}$ ($l=1,2,3$) denote the set of momentum variables of the inhomogeneous sector, canonically conjugate to the configuration variables $\{V^{\vec n,\epsilon}_{q_l}\}$. Here $\pi_{v_{\vec n,\epsilon}}$ are the momenta canonically conjugate to the modes $V^{\vec n,\epsilon}_{q_1}\equiv v_{\vec n,\epsilon}$ of the MS gauge-invariant, $\mathcal C^{\vec n,\epsilon}_{|1}$ denote the linear perturbative constraints that arise from perturbing (and abelianizing) the Hamiltonian constraint, and $\mathcal C^{\vec n,\epsilon}_{\_1}$ denote the linear perturbative constraints that arise from perturbing the diffeomorphisms constraint. These two sets of variables coordinatize the inhomogeneous sector of the phase space. The homogeneous sector describes the global mode of the scalar field, $\phi$, and that of the geometry, together with their canonically conjugate momenta. In order to later apply the loop quantization to the homogeneous gravitational sector, we choose the volume variable $v$, already introduced in previous sections, to capture the information about this zero-mode of the homogeneous geometry.\footnote{We may obtain the flat FRW geometry from the Bianchi I model of previous sections by identifying the triad and the connection variables in all directions. Then, $|v|\propto \sqrt{|p_\theta p_\sigma p_\delta|}$.} Furthermore we introduce some functions that depend exclusively on the above zero-modes: $\mathcal{H}_0^{(2)}$, $\vartheta_o$, $\vartheta_e$, and $\vartheta_e^{q}$, whose explicit expressions can be checked in Ref. \refcite{hyb-pert5}. These functions do not depend on the momentum canonically conjugate to the zero-mode of the scalar field, that we denote by $p_\phi$, nor on the eigenvalues $\omega_n^2$. We also define the following objects quadratic in perturbations:
\begin{align}\label{Thetas}
\Theta_o&\equiv \sum_{\vec n, \epsilon}- \vartheta_o v_{\vec n,\epsilon}^2\;,\qquad \Theta_e\equiv \sum_{\vec n, \epsilon}-\left[(\vartheta_e \omega_n^2+\vartheta_e^q)v_{\vec n,\epsilon}^2+ \vartheta_e \pi_{v_{\vec n,\epsilon}^2}\right]\;.
\end{align}
Then, the global mode of the Hamiltonian constraint of the perturbed system reads
\begin{align}\label{dens-constraint1}
\mathcal C=\frac12 \left[{p}_{\phi}^2-{\mathcal H}_0^{(2)}-{\Theta}_e-\Theta_o{p}_{\phi} \right]\;.
\end{align}
The first two terms of $\mathcal C$ form the Hamiltonian constraint of the unperturbed FRW model. Recalling Eq. \eqref{cfrw}, we see that $\mathcal{H}_0^{(2)}=3\pi G\hbar^2 \Omega^2/8$ if the field were massless, with $\Omega=cp/(\pi G\hbar \gamma)$ the classical counterpart of the LQC operator $\widehat{\Omega}$. In the case of a field subject to a potential, the potential contributes to the expression of $\mathcal{H}_0^{(2)}$.

We then adopt the hybrid approach to quantize the model, combining the LQC representation of the homogeneous sector \cite{aps,mmo} with a Fock quantization for the gauge-invariant perturbations. As we did for the Gowdy model, we need to deal with the ambiguities in the choice of the Fock representation. As mentioned in the introduction, for cosmological perturbations the criteria of vacuum invariance under the spatial isometries of the unperturbed FRW model and of unitarity of the quantum dynamics, select as well a unique equivalence class of Fock representations (see, e.g., Ref. \refcite{uni8}). We thus employ an element of that class to represent the gauge-invariant perturbations in our hybrid quantization. This provides a well-defined symmetric operator, 
\begin{align}\label{dens-constraint2}
\widehat{\mathcal C}=\frac12 \left[\hat{p}_{\phi}^2-\widehat{\mathcal H}_0^{(2)}-\widehat{\Theta}_e-\frac12\left(\widehat{\Theta}_o\hat{p}_{\phi}+\hat{p}_{\phi}\widehat{\Theta}_o\right) \right]\;,
\end{align}
corresponding to the zero-mode of the Hamiltonian constraint. Besides, the linear perturbative constraints are represented as derivatives (or as translations, in an integrated version of them). All the constraint operators commute, respecting that their classical algebra is abelian. Imposition of these linear perturbative constraints simply results in that physical states $\Psi$ do not depend on the inhomogeneous configuration variables  $\{V^{\vec n,\epsilon}_{q_2}, V^{\vec n,\epsilon}_{q_3}\}$, conjugate to those constraints.

Just as in the case of the Gowdy model, solving the remaining constraint, the global Hamiltonian constraint, is a difficult endeavour, and approximations are needed. A particularly interesting regime of the theory is that in which one can separate (but not necessarily decouple) the dynamics of the perturbations from the dynamics of the homogeneous FRW geometry. In order to study that regime, one adopts a Born-Oppenheimer ansatz of the form
\begin{equation}\label{BOans}
\Psi=\Gamma(v,\phi) \psi({\mathcal N},\phi),
\end{equation}
where the dependence on the MS variables is denoted by the label ${\mathcal N}$ of the  occupancy-number states for the gauge-invariant MS field. In this ansatz, the wave function $\Gamma(v,\phi) $ is normalized in the homogeneous FRW geometry, and evolves unitarily in $\phi$, that can be regarded as an internal time. We will call $\hat{U}$ the unitary operator (independent of ${\hat p}_{\phi}$) that provides this evolution. Moreover, we suppose that ${\hat U}$ can be chosen so that its difference with respect to the evolution of the geometry in the unperturbed case can be treated as small.\cite{hyb-pert5} For later convenience, we also define ${\hat h}\equiv[{\hat p}_{\phi},{\hat U}]{\hat U}^{-1}$.

Then, assuming that,  1) one can disregard changes in the FRW state $\Gamma$ mediated by the constraint,  2)  the expectation value on $\Gamma$ of $\widehat{\Theta}_{o}$ calculated with respect to the FRW geometry, $\langle \widehat{\Theta}_{o} \rangle_{\Gamma}$, is negligible as compared to $\langle {\hat h} \rangle_{\Gamma}$, and 3) the contribution of $\hat{p}_{\phi}^2 \psi$ is negligible, we obtain a Schr\"odinger-like equation for the quantum evolution of the perturbations, where $\phi$ plays the role of time \cite{hyb-pert5}. In this manner, we get a framework in which perturbations can be regarded as a test quantum field propagating on a quantum (mechanically corrected) background. This framework extends the quantum dynamics of perturbations beyond the onset of inflation. It is often called {\it dressed metric approach} and, with some further simplifications and assumptions, has been used to compute modifications to the standard predictions about the power spectrum of primordial scalar perturbations \cite{dres1,dres2,dres3,dres4,dres5}.

Moreover, under just our asumption 1 above (but not 2 and 3), and replacing quantum operators for the perturbations with their classical counterparts, that become (time-dependent) harmonic oscillators, it is easy to derive the effective dynamics for the gauge-invariant variables. One ends with the modifed MS equations \cite{hyb-pert5}
\begin{equation}
    d^2_{\eta_{\Gamma}}v_{\vec n,\pm} =- 2 \pi^2 v_{\vec n,\pm}  \left(2  \omega_n^2 + \frac{   \langle 2 {\hat \vartheta}_{e}^q  + ({\hat \vartheta}_{o}{\hat h} + {\hat h} {\hat \vartheta}_{o})+ [{\hat p}_{\phi}- {\hat h}, {\hat \vartheta}_{o}] \rangle_{\Gamma} } {  \langle   {\hat \vartheta}_{e}    \rangle_{\Gamma} } \right) .\end{equation}
Here, $\eta_{\Gamma}$ is a state-dependent conformal time, that takes into account the expectation value of the homogeneous volume on $\Gamma$.
The last term in our modified MS equations includes quantum corrections and, remarkably, is mode independent. Owing to this fact, we see that the equations remain of hyperbolic type in the ultraviolet regime. As above with the Schr\"odinger equation, these equations provide master equations to extract physical consequences of quantum gravity in cosmology.

\section{Conclusions}
\label{conclu}

This paper reviews the main ideas and constructions underlying the hybrid approach carried out in LQC to quantize inhomogeneous models. This approach assumes a physical regime where the global zero-modes of the geometry encode the main quantum gravity effects, and then the inhomogeneities, even having a quantum nature, can be treated in a more conventional way employing standard quantum field theory methods. 

The hybrid approach provides a well-defined framework for a regime in between the full loop quantum gravity regime and the regime of quantum field theory in classical spacetimes. In the particular case of the $T^3$--Gowdy model with linear polarization, discussed in Sec. \ref{gowdy}, the physical Hilbert space for the inhomogeneities, $ \mathcal{F}_{\text{p}}$, is actually equivalent to that obtained for them in the standard quantization of the deparameterized model.  Therefore we recover the standard description of the inhomogeneities. In the particular case of the scalar cosmological perturbations discussed in Sec. \ref{perturbations}, a Born-Oppenheimer approximation reduces the hybrid quantum dynamics to the framework of quantum field theory in {\it quantum (mechanically modified) spacetimes}, which in turn reduces to quantum field theory in classical spacetimes when the quantum effects on the homogeneous geometry are negligible. In summary, the hybrid approach provides a suitable and consistent framework to measure the main effects of quantum gravity in cosmology.

One major point in the hybrid quantization is the choice of Fock representation for the inhomogeneous sector. For the case of free scalar fields propagating in isotropic spacetimes, with compact spatial sections with dimension $d\leq 3$, there are uniqueness theorems fixing the ambiguity in the Fock representation \cite{uni0,uni1}. Indeed, the criteria of imposing invariance of the vacuum under the isometries of the spatial sections of the homogeneous sector, together with demanding a unitary implementability of the dynamics in the quantum theory, select a unique equivalence class of Fock representations. The condition of unitary dynamics guarantees that, during a finite period of time, the vacuum undergoes only a finite creation of particles. This condition is the natural relaxation of time-reparameterization invariance, imposed in stationary situations, when stationarity is no longer a symmetry, as it happens to be the case in cosmological spacetimes. 

Even though we focused our attention here on a hybrid approach that combines the representation of LQC for the homogenous sector with a Fock representation for the inhomogeneities, the hybrid approach can be generalized to other representations. For instance, one might consider other quantum-gravity inspired representations for the homogeneous geometry, different from that of LQC. Moreover, the strategy of employing loop quantization techniques for the geometry and a Fock quantization for fields propagating on that geometry is not limited to cosmology, and has been also carried out, for instance, in the context of black holes \cite{javi}.

\section*{Acknowledgments}

This work was partially supported by the MICINN/ MINECO Projects No. FIS2011-30145-C03-02 and FIS2014-54800-C2-2-P, and by The Netherlands Organisation for Scientific Research (NWO).


\begin{thebibliography}{99}   

\bibitem{boj1a} M. Bojowald,  {\it Class. Quantum Grav.}
{\bf17} (2000) 1489.

\bibitem{boj1b} M. Bojowald, {\it Class. Quantum Grav.}
{\bf17} (2000) 1509 .

\bibitem{boj1c} M. Bojowald, {\it Class. Quantum Grav.} {\bf18} (2001) 1055.

\bibitem{boj1d} M. Bojowald, {\it Class. Quantum Grav.} {\bf18} (2001) 1701. 

\bibitem{boj2} M. Bojowald, {\it Phys.\ Rev.\
Lett.} {\bf86} (2001) 5227.

\bibitem{abl} A. Ashtekar, M. Bojowald, and J. Lewandowski, {\it Adv.\ Theor.\ Math.\ Phys.} {\bf 7} (2003) 233.

\bibitem{aps} A. Ashtekar, T. Paw{\l}owski, and P. Singh, {\it Phys.\ Rev. D} {\bf 74} (2006) 084003.

\bibitem{acs} A. Ashtekar, A. Corichi, and P. Singh, {\it Phys.\ Rev. D} {\bf 77} (2008) 024046.

\bibitem{lqc1} M. Bojowald, {\it Living Rev.\ Rel.} {\bf 11} (2008) 4.

\bibitem{lqc2} G.A. Mena Marug\'an, {\it AIP Conf. Proc.} {\bf 1130}  (2009) 89.

\bibitem{lqc3} A. Ashtekar and P. Singh, {\it Class. Quantum Grav.} {\bf 28} (2011) 213001.

\bibitem{lqc4} K. Banerjee, G. Calcagni, and M. Mart\'in-Benito, {\it SIGMA} {\bf 8} (2012) 016.

\bibitem{hybrid-Gow1}
M. Mart\'in-Benito, L.J. Garay, and G.A. Mena Marug\'an, {\it Phys. Rev. D} {\bf 78} (2008) 083516.

\bibitem{hybrid-Gow2}
G.A. Mena Marug\'an and M. Mart\'in-Benito, {\it Int. J. Mod. Phys. A} {\bf 24} (2009) 2820. 

\bibitem{hybrid-Gow3}
L.J. Garay, M. Mart\'in-Benito, and G.A. Mena Marug\'an, {\it Phys. Rev. D} {\bf 82} (2010) 044048.

\bibitem{hybrid-Gow4}
M. Mart\'in-Benito, G.A. Mena Marug\'an, and E. Wilson-Ewing, {\it Phys. Rev. D} {\bf 82} 
(2010) 084012.

\bibitem{gowdy1} R.H. Gowdy, {\it Phys. Rev. Lett.} {\bf 27} (1971) 826.

\bibitem{gowdy2} R.H. Gowdy,  {\it Ann. Phys.} {\bf83}
(1974) 203.

\bibitem{awe} A. Ashtekar and E. Wilson-Ewing, 
 {\it Phys.\ Rev. D} {\bf 79} (2009) 083535.

\bibitem{men0}  J. Cortez and G.A. Mena Marug\'{a}n, {\it Phys. Rev. D} {\bf 72}  (2005) 064020.

\bibitem{men1a}
A. Corichi, J. Cortez, and G.A. Mena Marug\'{a}n, {\it Phys. Rev. D} {\bf 73} (2006) 041502.

\bibitem{men1b}
A. Corichi, J. Cortez, and G.A. Mena Marug\'{a}n, {\it Phys. Rev. D} {\bf 73}  (2006) 084020.

\bibitem{men2} A. Corichi, J. Cortez, G.A. Mena Marug\'{a}n, and J.M Velhinho, {\it Class. Quantum Grav.} {\bf23} (2006) 6301.

\bibitem{men3}  J. Cortez, G.A. Mena Marug\'{a}n, and J.M Velhinho, {\it Phys. Rev. D} {\bf 75}  (2007) 084027.

\bibitem{gow-matter}  M. Mart\'{\i}n-Benito, D. Mart\'{\i}n-de Blas, and G.A. Mena Marug\'{a}n,
 {\it Phys. Rev. D} {\bf 83} (2011) 084050.
 

\bibitem{barbero}  J.F. Barbero G., D. G\'omez Vergel, and E.J.S. Villase\~nor, {\it Class. Quantum Grav.} {\bf 24} (2007) 5945.

\bibitem{approx1}  M. Mart\'{\i}n-Benito, D. Mart\'{\i}n-de Blas, and G.A. Mena Marug\'{a}n,
{\it Class. Quantum Grav.} {\bf 32} (2014) 075022.

\bibitem{approx2} B. Elizaga Navascu\'es, M. Mart\'{\i}n-Benito, and G.A. Mena Marug\'{a}n,
 {\it Phys. Rev. D} {\bf 91} (2015) 024028.
 
\bibitem{approx3} B. Elizaga Navascu\'es, M. Mart\'{\i}n-Benito, and G.A. Mena Marug\'{a}n,
{\it Phys. Rev. D} {\bf 92} (2015) 024007.

\bibitem{hyb-pert1}
M. Fern\'andez-M\'endez, G.A. Mena Marug\'an, and J. Olmedo, {\it Phys. Rev. D}
{\bf 86} (2012) 024003.

\bibitem{hyb-pert2}
M. Fern\'andez-M\'endez, G.A. Mena Marug\'an, and J. Olmedo, {\it Phys. Rev. D}
{\bf 88} (2013) 044013.

\bibitem{hyb-pert3}
M. Fern\'andez-M\'endez, G.A. Mena Marug\'an, and J. Olmedo, {\it Phys. Rev. D}
{\bf 89} (2014) 044041.

\bibitem{hyb-pert4}
L. Castell\'o Gomar, M. Fern\'andez-M\'endez, G.A. Mena Marug\'an, and J. Olmedo, {\it Phys. Rev. D} {\bf 90} (2014) 064015.

\bibitem{hyb-pert5}
L. Castell\'o Gomar,  M. Mart\'{\i}n-Benito, and G.A. Mena Marug\'{a}n, {\it JCAP} 1506 (2015) 045. 

\bibitem{mmo} M. Mart\'in-Benito, G.A. Mena Marug\'an, and J. Olmedo, {\it Phys. Rev. D} {\bf 80} (2009) 104015.

\bibitem{mukhanov} V. Mukhanov, {\it Physical foundations of cosmology}, (Cambrige University Press, Cambridge, U.K., 2005).

\bibitem{uni0} J. Cortez, G.A. Mena Marug\'an, J. Olmedo, and J.M. Velhinho, {\it Phys. Rev. D} {\bf 83} (2011) 025002.

\bibitem{uni1} J. Cortez, G.A. Mena Marug\'an, J. Olmedo, and J.M. Velhinho, {\it Class. Quantum Grav.}  {\bf 28} (2011) 172001.

\bibitem{uni2} M. Fern\'andez-M\'endez, G.A. Mena Marug\'an, J. Olmedo, and J.M. Velhinho, {\it Phys. Rev. D} {\bf 85} (2012) 103525.

\bibitem{uni3} J. Cortez, G.A. Mena Marug\'an, J. Olmedo, and J.M. Velhinho, {\it Phys. Rev. D} {\bf 86} (2012) 104003.

\bibitem{uni4} L. Castell\'o Gomar, J. Cortez, D. Mart\'in-de Blas, G.A. Mena Marug\'an, and J.M. Velhinho, {\it JCAP} 1211 (2012) 001.

\bibitem{uni7} L. Castell\'o Gomar and G.A. Mena Marug\'an, {\it Phys. Rev. D} {\bf 89} (2014) 084052.

\bibitem{uni8}  J. Cortez, G.A. Mena Marug\'an, and J.M. Velhinho, {\it Ann. Phys.} {\bf 363} (2015) 36.

\bibitem{dres1}  I. Agullo, A. Ashtekar, and W. Nelson, {\it Phys. Rev. Lett.} {\bf 109} (2012) 251301.

\bibitem{dres2}  I. Agullo, A. Ashtekar, and W. Nelson, {\it Phys. Rev. D} {\bf 87} (2013) 043507.

\bibitem{dres3}  I. Agullo, A. Ashtekar, and W. Nelson, {\it Class. Quantum Grav.} {\bf 30} (2013) 085014.

\bibitem{dres4} I. Agullo, {\it Phys. Rev. D} {\bf 92} (2015) 064038.

\bibitem{dres5} I. Agullo and N.A. Morris, Detailed analysis of the predictions of loop quantum cosmology for the primordial power spectra, {\it  arXiv:1509.05693}.

\bibitem{precision1} G. Hinshaw {\it et al.}, {\it ApJS} {\bf 208} (2013) 19.

\bibitem{precision2} P.A.R. Ade {\it et al.} (Planck Collaboration), Planck 2015 results. XIII. Cosmological parameters, {\it arXiv:1502.01589}.

\bibitem{HH} J.J. Halliwell and S.W. Hawking, {\it Phys. Rev. D} {\bf 31} (1985) 1777.

\bibitem{javi} R. Gambini, J. Olmedo, and J. Pullin, {\it Class. Quantum Grav.} {\bf 32} (2015) 115002.

\end{thebibliography}
\end{document}